\documentclass{article}
\usepackage{epsfig}
\usepackage{bbm}
\usepackage{graphicx}
\usepackage{amsmath,amssymb}

\usepackage{booktabs}
\usepackage{textcomp}
\usepackage{makeidx}
\usepackage[bookmarks,colorlinks,breaklinks]{hyperref}
\newcommand{\comment}[1]{}

\newcommand{\EEA}{\end{eqnarray}}
\newcommand{\BEA}{\begin{eqnarray}}

\newcommand{\peq}{p_{eq}}
\newenvironment{example}[1][Example]{\begin{trivlist}
\item[\hskip \labelsep {\bfseries #1}]}{\end{trivlist}}

\makeatletter
\newcommand{\xleftrightarrow}[2][]{\ext@arrow 3359\leftrightarrowfill@{#1}{#2}}
\newcommand{\xdashrightarrow}[2][]{\ext@arrow 0359\rightarrowfill@@{#1}{#2}}
\newcommand{\xdashleftarrow}[2][]{\ext@arrow 3095\leftarrowfill@@{#1}{#2}}
\newcommand{\xdashleftrightarrow}[2][]{\ext@arrow 3359\leftrightarrowfill@@{#1}{#2}}
\def\rightarrowfill@@{\arrowfill@@\relax\relbar\rightarrow}
\def\leftarrowfill@@{\arrowfill@@\leftarrow\relbar\relax}
\def\leftrightarrowfill@@{\arrowfill@@\leftarrow\relbar\rightarrow}
\def\arrowfill@@#1#2#3#4{%
  $\m@th\thickmuskip0mu\medmuskip\thickmuskip\thinmuskip\thickmuskip
   \relax#4#1
   \xleaders\hbox{$#4#2$}\hfill
   #3$%
}
\makeatother

\begin{document}
\title{Data-Driven Computational Methods: Parameter and Operator Estimations (Chapter~1\footnote{This introductory chapter is taken from a book that will be published by Cambridge University Press (\url{http://www.cambridge.org/9781108472470}) and that it is in copyright.}~)}
\author{John Harlim\footnote{email: jharlim@psu.edu.} \\ Department of Mathematics and Department of Meteorology \\ The Pennsylvania State University}
\date{\today}
\maketitle

\begin{abstract}
Modern scientific computational methods are undergoing a transformative change; big
data and statistical learning methods now have the potential to outperform the classical
first-principles modeling paradigm. This book bridges this transition, connecting the
theory of probability, stochastic processes, functional analysis, numerical analysis, and
differential geometry. It describes two classes of computational methods to leverage
data for modeling dynamical systems. The first is concerned with data fitting algorithms to estimate parameters in parametric models that are postulated on the basis of physical or dynamical laws. The second class is on operator estimation, which uses the data to nonparametrically approximate the operator generated by the transition function of the underlying dynamical systems.

This self-contained book is suitable for graduate studies in applied mathematics,
statistics, and engineering. Carefully chosen elementary examples with supplementary
MATLAB$^{\mbox{\textregistered}}$ codes and appendices covering the relevant prerequisite materials are provided,
making it suitable for self-study.
\end{abstract}

\section{Introduction}

In applied science and engineering applications, modeling effort requires both physical insight in order to choose the appropriate mathematical models and computational tools for parameter inference and model validation.  Insofar as the physical intuition is concerned, one usually proposes a mathematical model based on certain physical law or observed mechanism. Unfortunately, the resulting models are typically subject to errors, be they of a systematic type due to incomplete physical understanding or of statistical nature due to uncertainties in the initial conditions, boundary conditions, model parameters, numerical discretization, etc. Since the ultimate goal of modeling dynamical systems is to predict the future states, it is important to compare the model-based predictions with the actual observables. It is also equally important to provide uncertainties associated with the predictions. As a consequence, the demand for computational methods that involve data fitting and uncertainty quantification is increasing. 

Traditionally, statistical science is the leading and established field that analyzes data and develops such computational tools. The focus of this book to a large extent is on surveying recent data-driven methods for modeling dynamical systems. In particular, we survey numerical methods that leverage observational data to estimate parameters in a dynamical model when the parametric model is available and to approximate the model nonparametrically when such parametric model is not available. These topics were developed through interaction between certain areas of mathematics and statistics such as probability, stochastic processes, numerical
analysis, spectral theory, applied differential geometry, Bayesian inference, Monte Carlo integrals, and kernel methods for density and operator estimations. Even with such a wide spectrum of interdisciplinary areas, the coverage here is far from complete. Nevertheless, we hope that the selected topics in this book can serve as a foundation for the data-driven methods in modeling stochastic dynamics.

\subsection{The role of data in parametric modeling}

Consider modeling dynamical systems in the form of differential equations,
\BEA
\frac{dx}{dt} = f(x,\theta),\label{ch1:ds}
\EEA
where $x(t;\theta)$ is the variable of interest and the vector field $f$ defines the ``law'' that determines how $x$ changes with time. Here, the differential equations can be either deterministic or stochastic. When the dependence of $f$ on state variables $x$ and parameters $\theta$ is given (or imposed), we call such a representation \index{parametric modeling}\emph{parametric} modeling. 

To make this dynamical model useful for predicting the future state, $x(t;\theta), t>t_i$, one needs to specify the parameters $\theta$ as well as initial conditions, $x(t_i)$, which reflects the current state. This inverse problem can naturally be solved with a Bayesian approach \cite{dashtistuart:2017}. This parameter estimation problem is the first main topic of this book. We will neglect the non-Bayesian approach in this book. 

Now, let us describe the basic idea of Bayesian approach. In practice, we often observe noisy discrete-time data, 
\BEA
y_i = h(x(t_i;\theta^\dagger,x_0)) + \eta_i,\label{ch1:obs}
\EEA
where the subscript $i$ denotes a discrete time index, $h$ denotes the observation operator, $x(t_i;\theta^\dagger,x_0)$ denotes the solutions of \eqref{ch1:ds} at time $t_i$ with hidden parameters $\theta^\dagger$ and initial condition $x(t_0)=x_0$. In \eqref{ch1:obs}, the terms $\eta_i$ denote unbiased independent and identically distributed (i.i.d.) noises, representing measurement error. Depending on the distribution of the observation error $\eta_i$, one can define the likelihood function of $(\theta,x_0)$ via the conditional distribution, $p(y_i|x_0,\theta) = p(y_i- h(x(t_i;\theta,x_0))) = p(\eta_i)$, where $x_0$ can also be estimated when it is not known. In this book, we will survey two popular Bayesian computational methods to estimate the conditional density for $\theta$ given the measured observations in \eqref{ch1:obs}.

\subsubsection{Markov-chain Monte Carlo}\index{MCMC}
Let's denote $y=\{y_1,\ldots,y_n\}$ and $p(y|\theta) = \prod_{i=1}^n p(y_i|\theta)$ and assume that the initial condition $x_0$ is given. The objective of this Bayesian inference is to estimate $p(\theta | y)$ by applying Bayes' rule, \index{Bayesian inference}
\BEA
p(\theta | y) \propto p(\theta) p(y|\theta),\label{ch1:bayes}
\EEA
where $p(\theta)$ denotes the prior density of the parameter. Here, the prior acts as a regularization term to overcome ill-posedness in the inverse problems \cite{dashtistuart:2017}. From the estimated posterior density $p(\theta|y)$, one can deduce statistical quantities, such as the mean as a point estimator for $\theta^\dagger$ and the covariance to quantify the uncertainty of the mean estimate. 

A popular method to sample the posterior density $p(\theta | y)$ in \eqref{ch1:bayes} is the Markov-chain Monte Carlo (MCMC), which will be discussed in Chapter~2. We will give a brief survey of the mathematical theory behind the MCMC to give readers a solid understanding of this sampling procedure. Briefly, this method constructs a Markov-chain with the posterior $p(\theta | y)$ as the limiting or target distribution. While the MCMC approach for solving the Bayes' formula in \eqref{ch1:bayes} is a ``gold standard'',  this objective is computationally demanding and may not be feasible when the underlying model in \eqref{ch1:ds} is high-dimensional. This computational overhead is because MCMC involves an iterative procedure that requires one to solve the dynamical system in \eqref{ch1:ds} on the proposed parameters in each iteration. One popular way to avoid this expensive calculation is with a surrogate modeling \cite{mx:09}, which will be discussed in Section~2.5. In Example~1.1, we give a brief illustration of the expected product of the MCMC implemented with the underlying dynamics as well as with a surrogate model constructed using a polynomial expansion (which detailed is discussed in Chapter~4).   

\begin{example}[Example 1.1]\label{ch1:ex1}\index{Lorenz-96 model}
Consider estimating two parameters $D,F$ of a system of a $5$-dimensional Lorenz-96 model \cite{lorenz:96},
\BEA
\frac{dx_j}{dt} &=& x_{j-1}(x_{j+1}-x_{j-2}) - D x_j + F,\quad\quad j=1,\ldots, J, \label{ch1:l96}\\
x_j(0) &=& sin(\frac{2\pi j}{5}) \nonumber
\EEA
from a given set of discrete-time observations, $\Delta t= t_i-t_{i-1}=0.05$, $i=1,\ldots, 10,$
\BEA
y_j(t_i) = x_j(t_i)  + \eta_{i}, \quad \eta_i\sim\mathcal{N}(0,0.01).\label{ch1:obs2}
\EEA
In \eqref{ch1:obs2}, the observations of state $x(t_i)$ are corrupted with i.i.d. Gaussian noises, $\eta_i$.

In Figure~\ref{ch1:figure1}, we show the resulting posterior density estimate from the MCMC with the underlying model in \eqref{ch1:l96} as well as with a surrogate modeling constructed using a polynomial expansion that avoids integrating the system of differential equations in \eqref{ch1:l96}. Notice that the true parameter values are within the posterior density estimates.

\begin{figure}
\centering\includegraphics[width=0.6\textwidth]{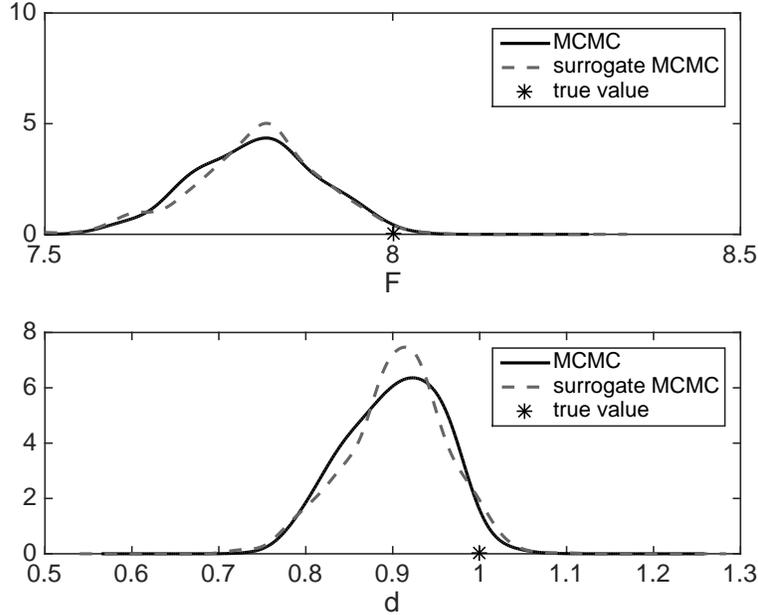}
\caption{Standard MCMC density of each parameter (black), surrogate MCMC density (dashes), and the true parameter value (asterisk).}
\label{ch1:figure1}
\end{figure}

\end{example}

\subsubsection{Ensemble Kalman filter}\index{ensemble Kalman filter}

The second Bayesian inference method we will discuss is the ensemble Kalman filter. In particular, define $\mathcal{Y}_i=\{y_j, j\leq i\}$. Here, we consider applying the Bayes' formula sequentially to approximate the posterior distribution of both the state and parameters, 
\BEA
p(\theta_i,x_i | \mathcal{Y}_{i}) \propto p(\theta_i, x_i | \mathcal{Y}_{i-1}) p(y_i | x_i,\theta_i)\label{ch1:bayes2}
\EEA
as the new observation $y_i$ becomes available. At each time step, we need to specify an initial density, $p(\theta_i,x_i | \mathcal{Y}_{i-1})$. 

Faithful solutions to the Bayesian filtering in \eqref{ch1:bayes2} have been proposed, such as the particle filter or sequential Monte Carlo \cite{ddfg:2001}, which represents the prior density with a point measure. However, clever sampling algorithms are needed to mitigate the curse of dimensionality of the classical particle filter \cite{bbl:08,blb:08}. In the world of applied science and engineering, a popular choice to approximate this Bayesian filtering problem is to use algorithms that are based on the celebrated Kalman filter \cite{kalman:61}. One of the most successful schemes that has been used in many applications, including numerical weather predictions, is the ensemble Kalman filter (EnKF) \cite{evensen:94}. The EnKF is a clever extension to the Kalman filter on nonlinear problems, without which the Kalman filter is impractical for high-dimensional problems. Since the Kalman filter formula is derived under strict assumptions, namely linearity and Gaussianity, it is clear that the EnKF, which represents the prior density, $p(\theta_i,x_i | \mathcal{Y}_{i-1})$, with a Gaussian measure, will not produce a meaningful estimate of the posterior density $p(\theta_i,x_i|\mathcal{Y}_i)$ in nonlinear and non-Gaussian problems.  Nevertheless, what is interesting is that often the ensemble solutions can track the true initial conditions, $x_i$, and parameter values, $\theta^\dagger$. In Chapter~3, we will discuss recent theoretical results that justify the accuracy of EnKF as a state estimation method, which has also been observed in many applications. The main emphasis of this chapter will be on the application of the EnKF in estimating both the state $x_i$ and the parameters $\theta$ in \eqref{ch1:ds}.  
In particular, we will focus on two parameter estimation methods. The first technique is a simple application of the EnKF to estimate parameters $\theta$ of the deterministic terms in the model. The second technique is on adaptive covariance estimation schemes that can be used in tandem with the EnKF to estimate parameters $\theta$ which represent the amplitudes of additive white noise forcings. In particular, we will discuss two recently developed methods that have been tested in many parameter estimation problems; the Berry-Sauer scheme \cite{bs:13} and the classical Belanger scheme \cite{belanger:74} which was recently adapted to EnKF \cite{hmm:14}. 

While fitting data to a dynamical model is a central topic of Chapters~2-3, constructing a model with accurate statistical prediction in the presence of model errors remains a challenging problem. In other words, constructing a model that can reproduce the marginal statistics (or observables) of the hidden dynamics is a nontrivial problem in general. In Section~3.3, we discuss this problem in the context of reduced-order modeling. Here, we survey the Mori-Zwanzig formalism \index{Mori-Zwanzig formalism}\cite{Mori65,Zwanzig73,Zwanzig61} as an idealistic concept for reduced-order modeling. Our goal with this discussion is to elucidate the difficulty of this problem. Subsequently, we will discuss a Markovian approximation for the generalized Langevin equation (GLE) derived from the Mori-Zwanzig formalism. \index{generalized Langevin equation (GLE)} Here, we will demonstrate the potential of using the parameter estimation scheme surveyed in this chapter to calibrate statistically accurate reduced-order Markovian dynamics. In Example~1.2, we give a brief illustration of the expected product from the parameter estimation method discussed in Chapter~3, implemented on a reduced-order model of a multiscale dynamical system.

\begin{example}[Example 1.2]\label{ch1:ex2}\index{Two-Layer Lorenz-96 model}
Consider the two-layer Lorenz-96 model \cite{lorenz:96}, whose governing equations are a system of $N(J+1)$-dimensional ODEs given by
\begin{equation}\label{ch1:lor96}
\begin{aligned}\frac{dx_i}{dt} &= x_{i-1}(x_{i+1}-x_{i-2}) - x_i + F + h_x\sum_{j=(i-1)J+1}^{iJ} y_{j}, \\
 \frac{dy_{j}}{dt} &= \frac{1}{\epsilon}\big(a y_{j+1}(y_{j-1}-y_{j+2}) - y_{j} + h_y x_{\textup{ceil}(i/J)} \big).
\end{aligned}
\end{equation}
Let $\vec{x}=(x_i)$ and $\vec{y}=(y_j)$ be vectors in $\mathbb{R}^{N}$ and $\mathbb{R}^{NJ}$ respectively, and the subscript $i$ is taken modulo $N$ and $j$ is taken modulo $NJ$. In this example, we set $N=8, J=32, \epsilon=.25, F=20, a=10, h_x = -0.4$, and $h_y=0.1$. In this regime the time scale separation is small. 

Suppose that we are given the following set of noisy observations:
\begin{align}
\vec{v}_m = h(\vec{x}(t_m)) + \eta_m, \quad \eta_m\sim\mathcal{N}(0,R), \nonumber
\end{align}
where $R=0.1\mathcal{I}_M$. In our experiment below, we will take observations only at every other grid point ($M=4$). That is, $h(\vec{x})=H\vec{x}$ is a linear observation function where $H\in\mathbb{R}^{4\times 8}$ and $H(i,2(i-1)+1)=1$ and zero everywhere else.

Consider the single-layer $N$-dimensional stochastically forced Lorenz-96 model \cite{bh:14} as the reduced-order model:
\BEA
\frac{d\tilde x_i}{dt} = \tilde x_{i-1}(\tilde x_{i+1}-\tilde x_{i-2}) -\tilde x_i + F - \alpha \tilde{x}_i(t) + \sigma \dot{W}_i(t),\label{ch1:reducedmodel}
\EEA
where $\dot{W}_i(t)$ denote white noises. Our goal is to estimate parameters $\alpha$ and $\sigma$ such that the reduced-order model in \eqref{ch1:reducedmodel} can reproduce the statistics of the slow components of \eqref{ch1:lor96}. Here, we estimate these parameters with an Ensemble Kalman Filter in tandem with the Berry-Sauer adaptive covariance method discussed in Chapter~3. For this example, see \cite{bh:14} for more detailed implementation and comparisons with other methods. In Figure~\ref{ch1:figure2}, we show the resulting statistical solutions of the reduced-order model in \eqref{ch1:reducedmodel} compared with the statistics of the true solutions of \eqref{ch1:lor96}. Notice that the reduced-order model is able to accurately reproduce the marginal density and autocorrelation function of the slow component, $x_i$, of  the full dynamics in \eqref{ch1:lor96}.  

\begin{figure}
\centering\includegraphics[width=0.48\textwidth]{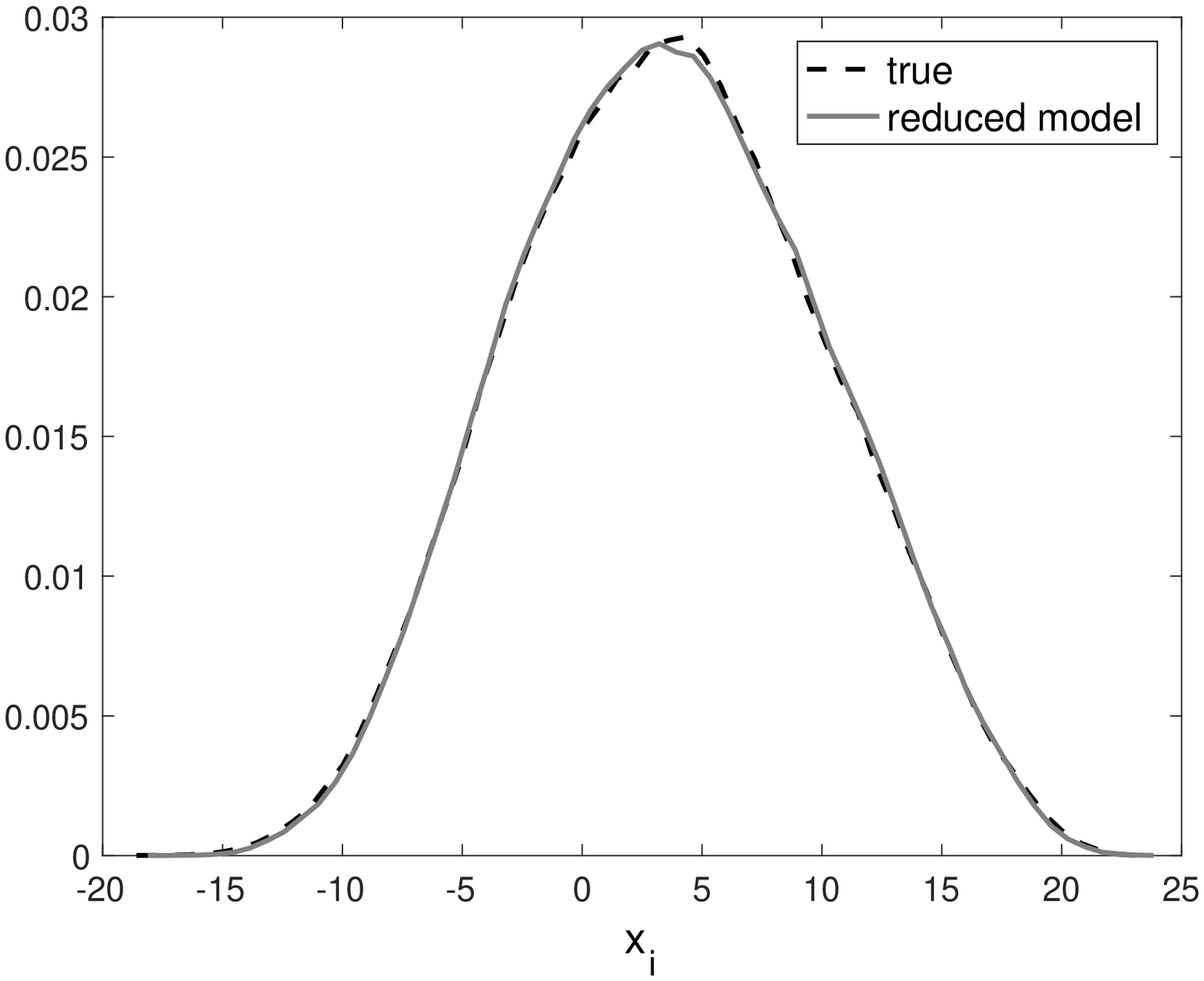}\quad\,\,
\centering\includegraphics[width=0.46\textwidth]{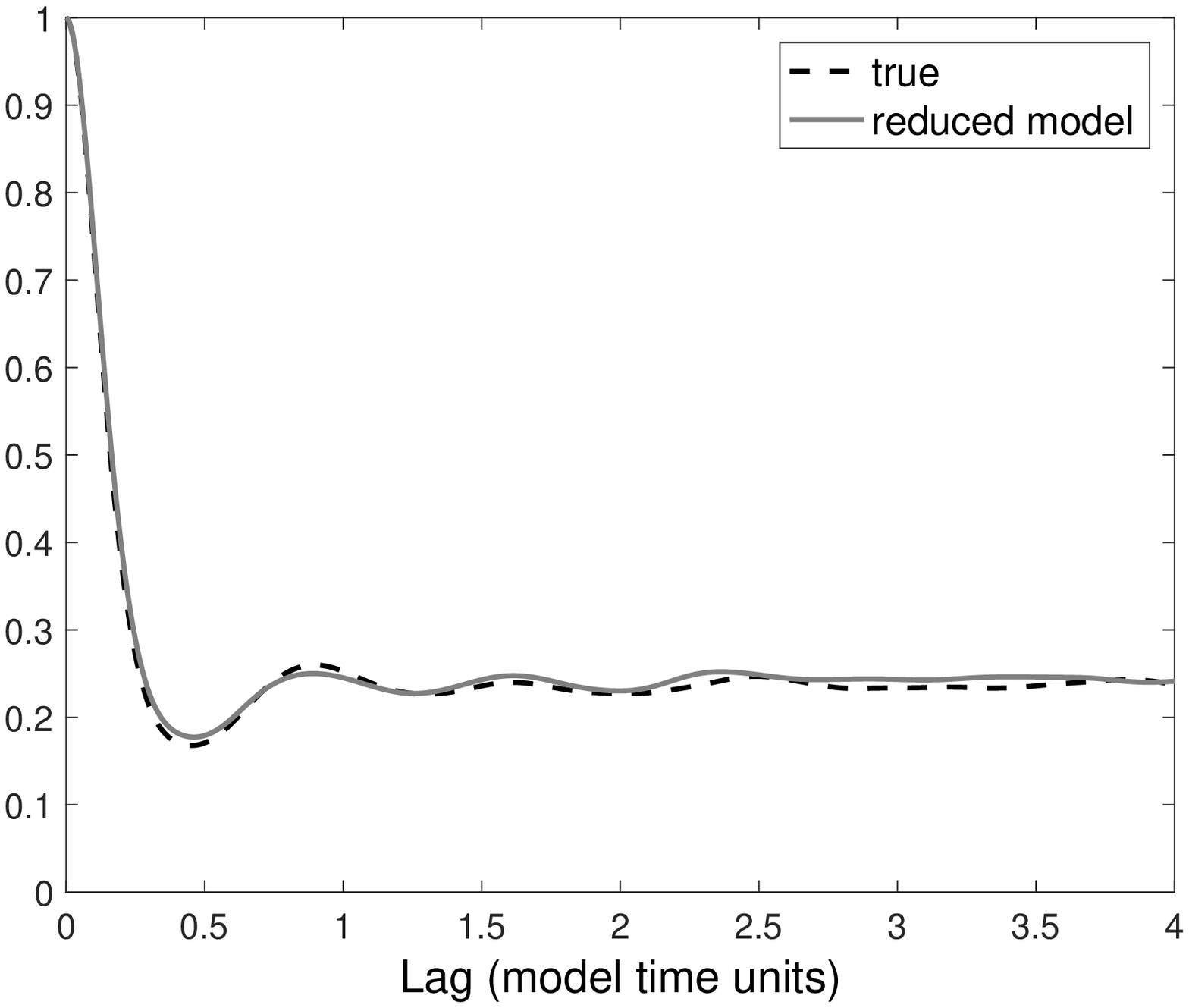}
\caption{Comparison of marginal density (left) and the time correlation (right) predicted by the reduced-order model in \eqref{ch1:reducedmodel} (grey solid) compared with the corresponding true statistics (black dashes) of \eqref{ch1:lor96}.}
\label{ch1:figure2}
\end{figure}

\end{example}

\subsection{Nonparametric modeling}

The second main topic of this book concerns an operator estimation method for \emph{nonparametric} modeling \index{nonparametric modeling} of dynamical systems. Our notion of nonparametric modeling follows directly from the standard statistical literature (see for example \cite{hardle2012}). That is, we do not make any strong assumption about how the vector field $f$ in \eqref{ch1:ds} depends on the state variables $x$ and parameters $\theta$. However, the method still contains parameters. For example, histogram is a nonparametric approach for estimating density functions and it contains parameters, namely the bin size and the number of bins. Kernel density estimation is another nonparametric approach for estimating density functions and it also has a parameter, namely the kernel bandwidth parameter. In fact, the kernel density estimate is usually implemented with a specific choice of kernel function, such as the Gaussian kernel, Epanechnikov kernel, etc. However, it is still considered nonparametric modeling in the sense that it does not make any a-priori assumption about the distributions that are being estimated, and the resulting estimate is independent with respect to the choice of kernel functions. In contrast, a parametric model for estimating densities imposes that the data be sampled from a certain distribution, such as the Gaussian, exponential, gamma, etc. 

An example of nonparametric modeling of dynamical systems is the \emph{analog} forecast, which finds states in the historical time series that are almost similar to the current state (it identifies analogs) and hopes that the history repeats itself \cite{lorenz:69}. Although this approach is less susceptible to model errors, it is difficult to identify the analog if the data space is high-dimensional even if the underlying dynamical systems are low-dimensional \cite{zg:14}. Furthermore, it is not so clear whether one can use this method for uncertainty quantification. 

In Chapter~6, we discuss a \emph{nonparametric probabilistic modeling} technique, the so-called \emph{diffusion forecast} \cite{bgh:15,bh:16physd}. This data-driven method rigorously approximates the solutions of the corresponding Fokker-Planck partial differential equations without knowing the differential operator. Since the solutions of the Fokker-Planck PDEs characterize the evolution of the distribution of the underlying dynamics, one can compute the corresponding time-dependent statistics to predict the future states and quantify uncertainties of the predictions. In a nutshell, the diffusion forecast is a method to solve a set of differential equations without knowing the equations. 

Our aim here is to show readers that the diffusion forecast is a natural extension of the central idea in uncertainty quantification (UQ), namely representation of random variables with a linear superposition of polynomial basis functions of appropriate Hilbert space. The main difference is that the diffusion forecast does not make any assumption on the distribution of the random variables which usually determines the polynomial basis functions as in the standard UQ. Instead it learns the basis functions from the data using a kernel-based nonlinear manifold learning method, the so-called \emph{diffusion maps} algorithm \cite{cl:06,bh:15vb}.\index{diffusion maps} We shall see that the diffusion forecast is a spectral Galerkin representation of the semigroup solution of the Fokker-Planck equation corresponding to the underlying dynamics with the data-driven basis functions.

With this intention, we include the following two related topics: the stochastic spectral method which has nothing to do with data and the Karhunen-Lo\`{e}ve expansion which applications include a linear manifold learning algorithm. Our main intention in including these two chapters is to demonstrate the transitional ideas passing from non-data-driven methods that are usually used in a parametric modeling context to a purely data-driven nonparametric modeling technique in the diffusion forecasting method.\index{Karhunen-Lo\`{e}ve} Readers who are familiar with these two topics can skip them and go directly to Chapter~6.

\subsubsection{Stochastic spectral method}

Given a parametric model as in \eqref{ch1:ds}, a popular subject known as Uncertainty Quantification (UQ) \cite{xiu:2010,mk:10} is concerned with estimating the following statistical quantities:
\BEA
\mathbb{E}[A(x)](t) = \int_{\cal M} A(x(t;\theta)) p(\theta)d\theta.\label{ch1:statistics}
\EEA
Here the parameters $\theta$ are assumed to be a realization of a random variable $\Theta$ with distribution $p(\theta)d\theta$ over the parameter domain $\cal M$. We also assume that $A\circ x \in L^1(\mathcal{M},p)$. The standard forward UQ technique imposes a certain assumption on the distribution of $\Theta$ and subsequently represents functions of $\Theta$ with a linear superposition of the orthogonal basis functions $\varphi_j(\theta)$ of the corresponding Hilbert space $L^2(\mathcal{M},p)$. For example, $\varphi_j(\theta)$ is the Hermite polynomial of degree-$j$ if $p$ is Gaussian or Legendre polynomial of degree-$j$ if $p$ is uniformly distributed. 

Given these basis functions, if $x$ is smooth as a function of $\theta$, one can approximate it as 
\BEA
x(t;\theta) \approx \sum_{k=1}^N x_k(t) \varphi_k(\theta), \label{ch1:pcexpansion} 
\EEA
where the time-dependent expansion coefficients,  $x_k(t) = \langle x(t;\cdot),\varphi_k \rangle_p$, are to be determined. With this approximation, one can estimate the integral in \eqref{ch1:statistics} for $A(x)=x^2$ as follows,
\BEA
\mathbb{E}[x^2](t) \approx \sum_{k,\ell=1}^N x_k(t) x_\ell(t) \int_{\cal M}  \varphi_k(\theta) \varphi_\ell(\theta) p(\theta)d\theta = \sum_{k=1}^N x^2_k(t),\nonumber
\EEA
thanks to the orthogonality property. In Chapter~4, we will discuss several approaches to compute the coefficients, $x_k(t)$, which may and may not involve deriving new equations based on the dynamics in \eqref{ch1:ds}.

As we mentioned before, this polynomial representation is non-data-driven. In fact, it chooses the basis functions by imposing certain assumptions on the distribution and the domain of the parameters. Since the representation idea is mathematically elegant, we would like to extend it non-parametrically. That is, our aim is to use the data to find the basis functions without making any assumption on the sampling distribution and the data manifold. Subsequently, we approximate smooth densities of the It\^o drifted diffusions with functions of a finite-dimensional subspace spanned by the resulting data-driven basis functions on $\mathcal{M}$. This is the central idea of the diffusion forecasting method. In contrast to the usual parametric approach, here we let the data determines the basis functions via the diffusion maps algorithm  \cite{cl:06,bh:15vb}. In fact, if the sampling measure of the data is Gaussian on the real line, then the resulting data-driven basis functions obtained via the diffusion maps algorithm are precisely the Hermite polynomials that are usually used in the orthogonal polynomial expansion for representing one-dimensional Gaussian random variable. Therefore, the data-driven basis that is used in the diffusion forecasting method is a natural generalization of the orthogonal polynomial basis on the data manifold.

\subsubsection{Karhunen-Lo\`{e}ve expansion} \index{Karhunen-Lo\`{e}ve}

The polynomial basis functions described in Chapter~4 can also be deduced from solving appropriate Sturm-Liouville eigenvalue problems with appropriate boundary conditions. This is an eigenvalue problem of a self adjoint second-order differential operator on a compact domain \cite{aa:1985}. This classical theory gives us an intuition behind the construction of the data-driven basis functions. Namely, our aim is to approximate a self-adjoint second-order differential operator on the compact manifold where the data lie, solve the corresponding eigenvalue problem, and set the resulting eigenvectors to be the discrete estimators of the basis functions. The diffusion maps algorithm \cite{cl:06} is a method that was designed to do these tasks on nonlinear data manifolds. 

To provide a self-contained exposition, we briefly review the Karhunen-Lo\`{e}ve expansion in Chapter~5. Our emphasis is to understand the Karhunen-Lo\`{e}ve expansion as an application of the Mercer's theorem that ties together the eigenfunctions of kernel-based integral operators and orthonormal basis functions of a Hilbert space. In an example, we will show that sometimes it is more convenient to transform the eigenvalue problem associated with an integral operator in Karhunen-Lo\`{e}ve expansion to an eigenvalue problem of a second-order elliptic differential operator. We shall see that the diffusion maps algorithm is exactly designed in the opposite way. This method is a kernel-based algorithm which approximates a weighted Laplacian operator on the data manifold with an integral operator. So, it approximates an eigenvalue problem of a differential operator by solving eigenvalue problem of an appropriate integral operator.  

While the basic theory of the KL expansion assumes the availability of the autocovariance function, one can also employ the KL expansion with an empirically estimated autocovariance function from the data. Intuitively, this approach represents the data in terms of the directions in which the data have the largest variance. The resulting method is a linear manifold learning algorithm which bears many names depending on the field of applications, including the Proper Orthogonal Decomposition (POD), Principal Component Analysis (PCA), Empirical Orthogonal Function (EOF), etc. By linear manifold learning, we refer to the fact that PCA represents data by a linear projection on a set of basis functions of a linear manifold, namely the ellipsoid. Specifically, the basis functions, which are usually called the principal components, are the axes of the ellipsoid. On the other hand, the diffusion maps algorithm (which will be discussed in Chapter~6 in detail) is a nonlinear manifold learning algorithm since it provides basis functions on an arbitrary data manifold embedded in a Euclidean space. \index{Principal Component Analysis (PCA)}

To clarify the distinction between linear and nonlinear manifold learning, we compare the principal components obtained from the POD and the basis functions obtained from the diffusion maps on a trivial yet illuminating example.  

\begin{example}[Example 1.3]\label{ch1:ex3a}
Consider uniformly sampled data, $x_i = (\cos(\theta_i), \sin(\theta_i))^\top$, $i=1,\ldots, N$, on a unit circle $S^1$ embedded in $\mathbb{R}^2$. Here, $\theta_i$ denotes the $i$th sample on the intrinsic coordinate of the circle, $S^1$. For clarity of exposition, in our numerical test on this artificial example below we generate ``very nice'' samples, with uniformly spaced $\theta_i = 2\pi i/N$. In practice, we usually don't have such a nice data set and the accuracy of the estimates will depend on the samples. Denote $X=[x_1, x_2,\ldots, x_N]\in\mathbb{R}^{2\times N}$. 

Loosely speaking, the goal of manifold learning is to find (basis) functions $\varphi(x)$ that can describe the data $x\in\mathcal{M}$. In particular, POD describes the data in terms of principal components which are defined as follows. The $k$th principal component (of POD) is defined as a functional $\psi_k(x)= w_k^\top x$ where $w_k$ solves the symmetric positive definite eigenvalue problem $\frac{1}{N}XX^\top w_k = \lambda_k w_k$. For this trivial circle example, $k=1, 2$, and
\BEA
\frac{1}{N}XX^\top \longrightarrow A = \begin{pmatrix} 1/2 & 0 \\ 0 & 1/2 \end{pmatrix},\nonumber
\EEA
as $N\to\infty$. In this case, the limit can be estimated analytically as follows 
\BEA
A_{ij} = \frac{1}{2\pi}\int _0^{2\pi} x^{i}(\theta) x^{j}(\theta)  d\theta = \frac{1}{2} \delta_{ij}.\nonumber
\EEA
Here, the notation $x^{j}(\theta)$ denotes the $j$th component of $x\in\mathbb{R}^2$, that is,
\BEA
x^j(\theta) = \begin{cases}\cos(\theta) & \mbox{if }j=1, \\ \sin(\theta) & \mbox{if }j=2. \end{cases}\nonumber
\EEA
Since the standard bases $e_1,e_2\in \mathbb{R}^2$ are eigenvectors of $A$, the principal components are nothing but $\psi_1(x) = e_1^\top x = x^1$ and $\psi_2(x) = e_2^\top x = x^2$. Essentially, each component of the given data (or each row of matrix $X$) is the principal component. To clarify this assertion, we plot the principal components (in color) as functions of the data in Figure~\ref{fourierexamplepc}. Notice that the first principal component identifies the data in the horizontal direction (the function values increase from -1 to 1). On the other hand, the second principal component identifies the data in the vertical direction. These two axes correspond to the principal axes of the unit circle. In general, the principal components of POD correspond to principal axes of an ellipsoid that is fitted to the data even if the data do not lie on an ellipsoid (see the example in Chapter~5).  

\begin{figure}
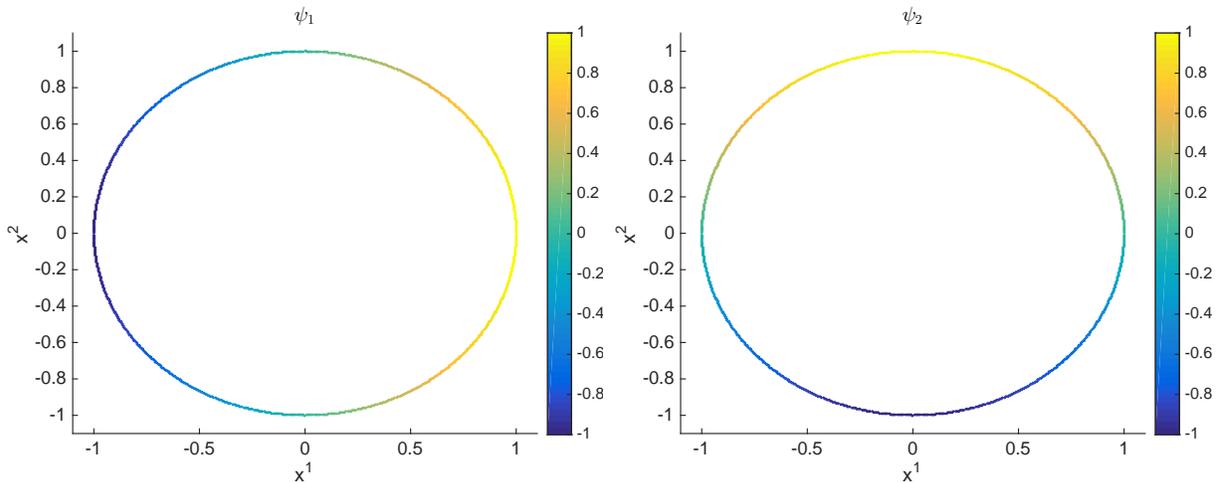

\centering
\includegraphics[width=0.48\textwidth]{fourierexamplepc1.eps}
\includegraphics[width=0.48\textwidth]{fourierexamplepc2.eps}
\caption{\label{fourierexamplepc} The principal components (color) as functions of the data.}
\end{figure}

On the other hand, the diffusion maps algorithm solves the following eigenvalue problem, $\Delta_\theta \varphi_k(\theta) = \lambda_k \varphi_k(\theta)$, where the Laplace-Beltrami operator is numerically estimated using a matrix as a discretization of a kernel-based integral operator. For this example, since the embedding function (or the Riemannian metric) is known, it is clear that the Laplace-Beltrami operator is simply a one-dimensional derivative with respect to the intrinsic coordinate. The explicit solutions of this eigenvalue problem are the Fourier series, $\varphi_k(\theta)=e^{ik\theta}$ which form a basis for $L^2(S^1)$. In Figure~\ref{fourierexample}, we compare the discrete estimates of the first four Fourier modes obtained from the diffusion maps algorithm applied on the data $\{x_i\}_{i=1,\ldots,N}$, where $N=1000$ with the corresponding analytical solutions. 

In summary, the diffusion maps algorithm produces orthonormal basis functions of a Hilbert space $L^2(S^1)$, where each component $\varphi_k:\mathcal{M}\to \mathbb{R}$ is a nonlinear map. In contrast, the principal components of POD are the first Fourier mode in this example, where each principal component is a linear function of the data manifold. 

\begin{figure}
\centering
\includegraphics[width=0.7\textwidth]{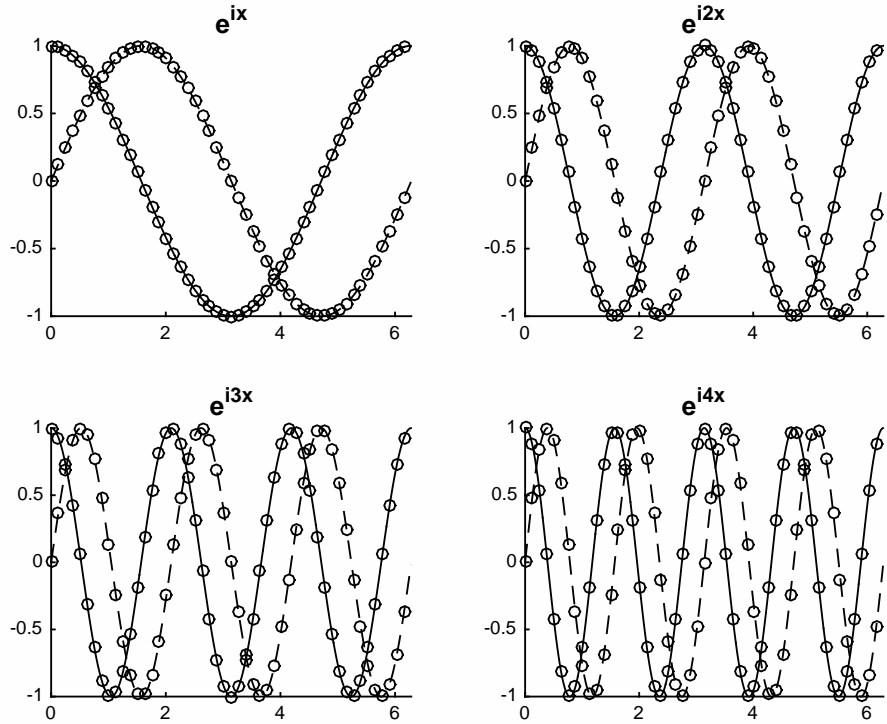}
\caption{\label{fourierexample} Discrete estimates of the eigenfunctions $e^{ikx}$, for $k=1,\ldots, 4$ evaluated on the training data manifold (circles) compared to the analytical solutions. In each panel, we show the cosine (solid) and sine (dashes) components. }
\end{figure}

\end{example}

In the next example, we show the application of the diffusion maps algorithm on a data set with a complicated manifold corresponding to solutions of a chaotic dynamical system. The key point is that we don't have the knowledge of the embedding (or the Riemannian metric) for the following example. As a consequence, we don't have an analytical expression for differential operators on this data manifold whose components are samples of the invariant measure of the dynamical systems. In this situation, diffusion maps algorithm is a powerful tool that approximates the weighted Laplacian operator on this complicated data manifold, where the weight is defined with respect to the sampling measure of the data.

\begin{example}\label{ch1:ex3}\index{Lorenz-63 model}
Consider the famous three-dimensional chaotic Lorenz model \cite{lorenz:63} which is a truncated approximation to the Navier-Stokes \index{Navier-Stokes} equations. This toy model is found to be useful to describe laser physics \cite{haken:75} and it is well-known as the first example of simple deterministic dynamical systems with solutions that are sensitive to initial conditions; this behavior has been called deterministic chaos or simply chaos. The governing equation of the Lorenz-63 model is given as
\begin{eqnarray}
\frac{dx}{dt} &=& \sigma(y-x) \nonumber \\ 
\frac{dy}{dt} &=& \rho x-y-xz \label{ch1:lorenz63} \\ 
\frac{dz}{dt} &=& xy-bz. \nonumber
\end{eqnarray}
with the parameter set $(\sigma,b,\rho)$, where in its original derivation \cite{lorenz:63,snm:96}, $\sigma$ is called the Prandl number and $\rho$ is the Rayleigh number.

In Figure~\ref{ch1:fig3}, we show the nonparametric estimates of the basis functions obtained via the diffusion maps algorithm, implemented with variable-bandwidth kernels (which will be discussed in Chapter~6). These eigenfunctions are generated using solutions $(x_i,y_i,z_i)$ of \eqref{ch1:lorenz63} at 5000 discrete time instances, with time step $t_{i+1}-t_i=\Delta t=0.5$ (see \cite{bgh:15} for the computational detail). In each of these panels, we depict the discrete estimate of the eigenfunction evaluated on each training data point, $\varphi_j(x_i,y_i,z_i)$. 

\begin{figure}
\centering
\includegraphics[width=0.7\textwidth]{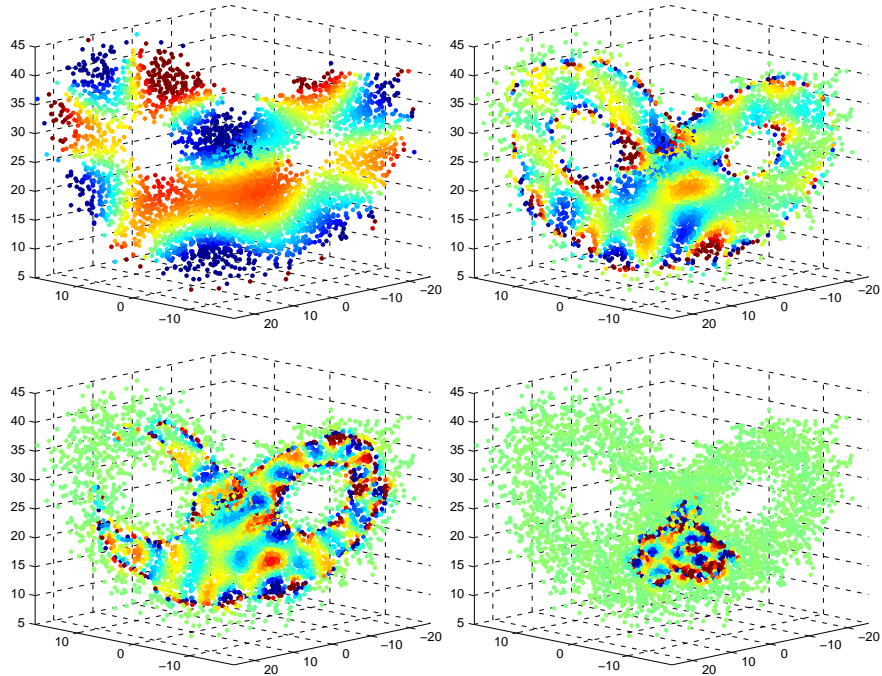}
\caption{\label{ch1:fig3} Discrete estimates of the eigenfunctions $\varphi_{40},\varphi_{500},\varphi_{1500},$ and $\varphi_{4000}$ evaluated on the training data manifold.}
\end{figure}

\end{example}

From these two examples, we can view the diffusion maps algorithm as a numerical method to estimate generalized Fourier basis (or orthogonal polynomials) of Hilbert space on the data manifold. Next, we will give a brief description of the diffusion forecasting method using these data-driven basis functions.
 
\subsubsection{Diffusion forecasting}\index{diffusion forecast}

Suppose that $x(t)\in\mathcal{M}\subset \mathbb{R}^n$ denotes a time-dependent It\^{o} diffusion, which satisfies a system of differential equations,
\BEA
dx = a(x)\,dt + b(x)\,dW_t,\label{ch1:dynsys}
\EEA
where $a(x)$ and $b(x)$ denote the drift and diffusion terms, respectively. Here, $dW_t$ denotes white noises. 

Assume that the model in \eqref{ch1:dynsys} is unknown, which means that the corresponding Fokker-Planck (or Liouville if \eqref{ch1:dynsys} is deterministic) equation, 
\BEA
\frac{\partial p}{\partial t} = \mathcal{L}^*p, \label{ch1:fp}
\EEA
is also unknown. Instead we are only given a set of time series $X = \{x_i$, $i=1,\ldots,N\}$ from measurements; here $x_i=x(t_i)$ are the solutions of \eqref{ch1:dynsys} given an initial condition $x_0$. We assume that $N$ is finite but large enough such that all configurations of the dynamics (or points in $\mathcal{M}$) are sufficiently close to some components in $X$. Given such practical constraints, the diffusion forecasting method uses these data to train a nonparametric probabilistic model whose solutions approximate the probability density function of $x$ at any time $t$. The key idea of this method is to represent an approximation of the semigroup solutions of the generator of \eqref{ch1:dynsys} with data-adapted basis functions $\varphi_j(x_i)$, obtained via the diffusion maps algorithm \cite{cl:06,bh:15vb}. In particular, we will represent the solutions of \eqref{ch1:fp} as follows,
\BEA 
p(x,t) = e^{t\mathcal{L}^*} p(x,0) = \sum_j \langle e^{t\mathcal{L}^*} p(x,0),  \varphi_j \rangle_{\peq} \varphi_j(x) \peq(x),
\EEA
where $\peq(x)$ denotes the equilibrium measure of the dynamical system in \eqref{ch1:dynsys}, such that $\mathcal{L}^*\peq=0$. Subsequently, we employ a nonparametric  approximation to the time-evolving coefficients $\langle e^{t\mathcal{L}^*} p(x,0),  \varphi_j \rangle_{\peq}$ such that we don't need to know $\mathcal{L}^*$. Since the diffusion map algorithm is a kernel-based method, one can interpret the diffusion forecast as an extension of the kernel density estimation method to estimate operators of Markovian dynamical systems. This is the topic of Chapter~6. In the next example, we give a brief illustration of the expected product of the diffusion forecast applied on time series of a chaotic dynamical system, the famous Lorenz-63 model.

\begin{example}[Example 1.5]
In Figure~\ref{ch1:fig4}, we show snapshots of the probability density at various times, obtained from the diffusion forecasting method, on the three-dimensional Lorenz-63 model in \eqref{ch1:lorenz63}. For qualitative comparison, we also show the Monte Carlo approximation of the evolution of the density (or ensemble forecasting), assuming that the full Lorenz-63 model is known. Here, the Monte Carlo initial conditions are prescribed by sampling the Gaussian density used in the diffusion forecasting method (as shown in the first row in Figure~\ref{ch1:fig4}). In each panel of this figure, we show the density as a function of $x+y$ and $z$ (corresponding to the three components of the Lorenz model) at different instances. In the left column, we also show the data set that is used for training the diffusion model (smaller dots). Notice that even at a long time $t=2$ (which is longer than the doubling time of this model, 0.78), the densities obtained from both forecasting methods are still in a good agreement. From these time-evolving density functions, one can compute statistical quantities for state estimation as well as uncertainty quantification nonparametrically.

\begin{figure}
\centering
\includegraphics[width=0.7\textwidth]{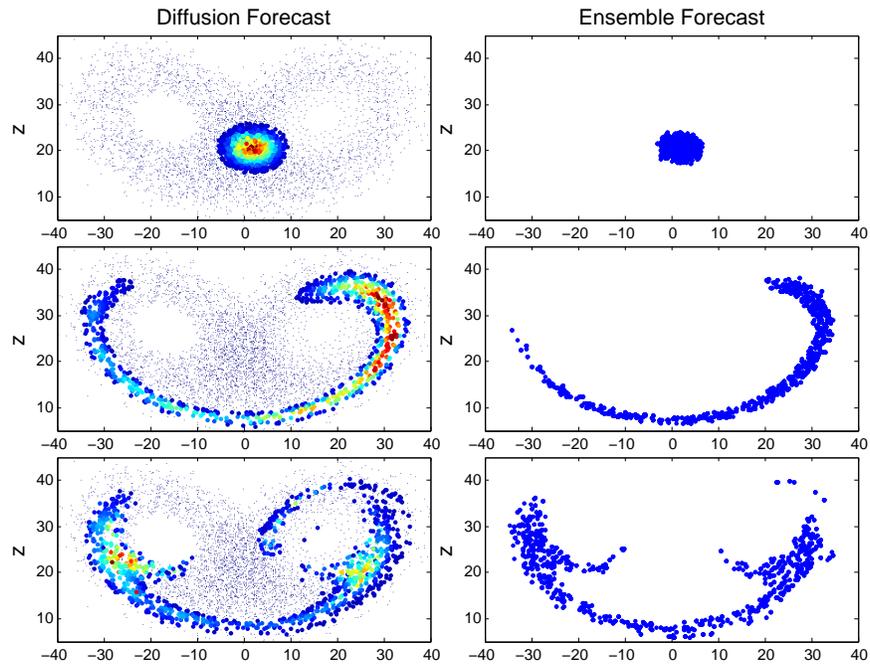}
\caption{Probability densities (as functions of $x+y$ and $z$) from the equation-free Diffusion Forecasting model (left column) and an ensemble forecasting (right column) at times $t=0$ (first row), $t=0.5$ (second row), and $t=2$ (third row). In the column on the left, the color spectrum ranging from red to blue is to denote high to low value of density.}
\label{ch1:fig4} 
\end{figure} 

\end{example}

While the example above assumes that the initial density is given, it is important to stress that in practice the initial density is not known. Usually, one is interested in predicting the future states with initial configurations, $x_j$, that are not in the training data set $X$. Therefore one needs to specify the corresponding initial densities, $p(x|x_j)$, to be used for predictions. In Chapter~6, we will also discuss the Nystr\"om extension and a Bayesian filtering method to specify these initial distributions for noiseless and noisy data, respectively.


\end{document}